\documentclass[twocolumn,superscriptaddress,floatfix,prb,amsmath,aps,showpacs]{revtex4}
\usepackage{graphicx}
\usepackage{subfigure}
\usepackage{bm}
\usepackage{amssymb}
\usepackage{url}
\usepackage{layout}
\usepackage{epsfig}
\usepackage{epstopdf}
\usepackage{booktabs}
\usepackage{float}
\usepackage{color}
\begin{document}
%\floatsep=1cm
\bibliographystyle{apsrev}
\def\nn{\nonumber}
\def\dag{\dagger}
\def\u{\uparrow}
\def\d{\downarrow}
\def\j{\bm j}
\def\m{\bm m}
\def\l{\bm l}
\def\0{\bm 0}
\def\k{\bm k}

%\begin{document}
\title{\textbf{Integer Quantum Magnon Hall Plateau-Plateau Transition in  a Spin Ice Model}}

\author{Baolong Xu}
\affiliation{International Center for Quantum Materials, Peking University, Beijing 100871, China}
\affiliation{Collaborative Innovation Center of Quantum Matter, Beijing 100871, China}
\author{Tomi Ohtsuki}
\affiliation{Department of Physics, Sophia University, Chiyoda-ku, Tokyo 102-8554, Japan}
\author{Ryuichi Shindou}
 \email{rshindou@pku.edu.cn}
\affiliation{International Center for Quantum Materials, Peking University, Beijing 100871, China}
\affiliation{Collaborative Innovation Center of Quantum Matter, Beijing 100871, China}
\date{\today}

\begin{abstract}
Low-energy magnon bands in a two-dimensional spin ice model
become integer quantum magnon Hall bands under an out-of-plane 
field. By calculating the localization length 
and the two-terminal conductance of magnon transport, we show that the
magnon bands with disorders undergo a quantum phase transition
from an integer quantum magnon Hall regime to a conventional magnon
localized regime. Finite size scaling analysis as well as a critical
conductance distribution shows that the quantum critical point
belongs to the same universality class as that in the quantum Hall transition.
We characterize thermal magnon Hall conductivity in disordered quantum magnon
Hall system in terms of robust chiral edge magnon transport.
\end{abstract}
\maketitle
Bosonic analogue of integer quantum Hall states have
been proposed in a number of quasi-particle boson systems with
broken time-reversal symmetry such as
photon,~\cite{haldane08,raghu08,wang09,ao09,ochiai09,lu14} phonon,~\cite{prodan09} exciton,~\cite{yuenzhou14}
exciton-polariton,~\cite{karzig14} triplon,~\cite{romhanyi15} magnon~\cite{shindou13a,shindou13b,shindou14,LifaZhang13,mook15,chisnell15,roldan-molina,kim16,owerre}
and surface magnon-polariton.~\cite{ochiai16} Typically, their quasi-particle excitations have
extended bulk bands with topological integers
and topological edge modes whose chiral dispersions
cross band gaps among these bulk bands.
Due to its chiral (unidirectional) nature, a quasi-particle boson flow along the
edge mode is believed to
be robust against generic elastic backward scatters, fostering
a rich prospect of their future applications.~\cite{raghu08,haldane08,wang09,ao09,ochiai09,lu14,prodan09,yuenzhou14,
karzig14,shindou13a,shindou13b,shindou14,ochiai16} 
On the one hand, these bosonic systems often {\it break
conservation of the quasi-particle number} even at the level of respective
quadratic Hamiltonian.~\cite{yuenzhou14,karzig14,shindou13a,shindou13b,shindou14,LifaZhang13,mook15,
chisnell15,kim16,engelhardt15a,furukawa15,galilo15,bardyn16,engelhardt15b,peano16,xu16} Thereby,
one naturally wonders if the quasi-particle 
flow along the topological edge modes is still
robust against such particle-number-non-conserving perturbations or not.
In other words, one may raise a question whether two quantum 
Hall regimes with different Chern integers are topologically distinguishable even
in the absence of the U(1) symmetry
associated with the quasi-particle number conservation.

In this rapid communication, 
we study effects of generic disorder potentials in a simplest spin model in a
quantum magnon Hall regime. Our numerical results and the following
argument clarify that, even without the explicit U(1) symmetry at the 
Hamiltonian level, the topological magnon edge mode provides
a robust quantized magnon conductance and therefore quantum magnon
Hall regimes with different topological integers are always distinguished by a quantum
critical point with delocalized  bulk magnon band.  Thermal conductance
distributions calculated at the critical point clearly shows 
that the quantum critical
point belongs to the same universality class as the
two-dimensional integer quantum Hall plateau-plateau
transition. Based on these knowledge,
we give a generic expression for the
thermal Hall conductivity in disordered integer quantum bosonic
Hall systems from edge transport picture.

We study spin excitations in a square-lattice
spin ice model~\cite{wang06,budrikis10,budrikis12,iacocca16} under out-of-plane Zeeman field $H_Z$;
\begin{align}
H= & \sum_{<{\bm i},{\bm j}>} \frac{1}{|{\bm i}-{\bm j}|^3}
(\hat{S}_{\bm i} \cdot \hat{S}_{\bm j}-3(\hat{S}_{\bm i}
\cdot {\bm n}_{{\bm i}{\bm j}})(\hat{S}_{\bm j}
\cdot {\bm n}_{{\bm i}{\bm j}})) \nn \\
& -D \sum_{{\bm i} \in A} {(\hat{S}^x_{\bm i})}^2 -
D \sum_{{\bm j} \in B} {(\hat{S}^y_{\bm j})}^2 -
H_Z \sum_{{\bm i} \in A,B}\hat{S}^z_{\bm i}. \label{spin-H}
\end{align}
The model consists of two inequivalent spins in a unit cell,
$A$-sublattice spin $S_{{\bm i} \in A}$ on the $x$-link of
the square lattice and $B$-sublattice spins $S_{{\bm i} \in B}$
on the $y$-link.~\cite{supple} Due to a magnetic shape
anisotropy,~\cite{wang06,budrikis10,budrikis12,iacocca16,supple} 
each sublattice spin has an easy-axis anisotropy $D$ $(>0)$ along
respective spatial direction. Heisenberg spins are coupled with
each other by magnetic dipole-dipole interaction,
${\bm n}_{{\bm i}{\bm j}}$ denotes the unit vector connecting 
sites ${\bm i}$ and ${\bm j}$. An inclusion of the
next and the next nearest neighbor magnetic dipolar couplings
imposes so-called two-in two-out ice rule for each vertex,
which has been experimentally observed in
a patterned ferromagnetic film.~\cite{wang06} 

When the classical ground-state spin configuration becomes fully polarized by
the Zeeman field ($H_Z > H_s \simeq DS$), the lowest magnon
band and the second lowest magnon band acquire the
topological number with opposite sign due to the finite next nearest
neighbor dipolar coupling, and a topological chiral edge
mode appears inside a band gap between the two.~\cite{supple} 
%The chirality of the edge mode against the the field direction is found to be right-handed.
%; same as the Damon-Eschbach surface mode.~\cite{damon61}
The corresponding magnon Hamiltonian is obtained
from Eq.~(\ref{spin-H})
with $S_{z,{\bm j}} \equiv S - b^{\dagger}_{\bm j} b_{\bm j}$,
$S_{-,{\bm j}}\equiv b^{\dagger}_{\bm j}\sqrt{2S}$, $S_{+,{\bm j}}\equiv
b_{\bm j}\sqrt{2S}$ as,
\begin{align}
{\bm H}_{b} \equiv & {\bm H}_{\rm on} + {\bm H}_{\rm nn} + {\bm H}_{\rm nnn} \nn \\
{\bm H}_{\rm on} \equiv & \frac{1}{2} \sum_{{\bm j} \in A} \big\{(D + d_{\bm j})S (-{b^{\dagger}_{{\bm j}}}^2
- b^{\dagger}_{\bm j} b_{\bm j}) + (H_Z + h_{\bm j}) b^{\dagger}_{\bm j} b_{\bm j}\big\}  \nn \\
& \hspace{-1.0cm} + \frac{1}{2} \sum_{{\bm j} \in B} \big\{(D + d_{\bm j}) S ({b^{\dagger}_{{\bm j}}}^2
- b^{\dagger}_{\bm j} b_{\bm j}) + (H_Z + h_{{\bm j}}) b^{\dagger}_{\bm j} b_{\bm j}\big\} + {\rm h.c.} \nn
\end{align}
\begin{align}
{\bm H}_{\rm nn} \equiv & \sum_{m=1,2} \sum_{{\bm j} \in A}
\sum_{{\bm i}={\bm j}\pm \delta_{m},{\bm i} \in B}  \nn \\
%\end{align}
%\begin{align}
& \hspace{-1.2cm}  \frac{J S}{4}
\big(  b^{\dagger}_{\bm j} b_{\bm i}   -2 b^{\dagger}_{\bm i}b_{\bm i}
 -2 b^{\dagger}_{\bm j} b_{\bm j} + 6i (-1)^m  b^{\dagger}_{\bm j} b^{\dagger}_{\bm i} + {\rm h.c.} \big) \nn \\
%\end{align}
%\begin{align}
{\bm H}_{\rm nnn} \equiv & \sum_{\alpha=A,B} \sum_{m=1,2}
\sum_{{\bm j} \in \alpha} \sum_{{\bm i}={\bm j}\pm e_{m}}  \nn \\
&\hspace{-1.2cm}  \frac{J^{\prime}_{\alpha,m} S}{4}
\big(  - b^{\dagger}_{\bm j} b_{\bm i} - b^{\dagger}_{\bm i}b_{\bm i}
 - b^{\dagger}_{\bm j} b_{\bm j} + 3 (-1)^m  b^{\dagger}_{\bm j} b^{\dagger}_{\bm i} + {\rm h.c.} \big) \label{boson-H}
\end{align}
where $J$, $J^{\prime}_{\alpha,m}$ denote the nearest and the next nearest dipolar interaction, 
respectively with $J^{\prime}_{A,1}=J^{\prime}_{B,2} = J^{\prime}_1$ and
$J^{\prime}_{A,2}=J^{\prime}_{B,1} = J^{\prime}_2$. $e_{1}$ and $e_{2}$ are the primitive lattice
vectors of the square lattice and $2\delta_{m}=e_1-(-1)^{m} e_2$ ($m=1,2$).~\cite{supple} 
Short-ranged randomness are introduced in Eq.~(\ref{boson-H}); $d_{\bm j}$ and
$h_{\bm j}$ are uniformly distributed within $[-W_D,W_D]$ and $[-W_H,W_H]$. We
set the unit of energy to be $JS$ and that of length the lattice spacing.
Due to magnetic anisotropy term, dipolar interaction
and randomness, the quadratic boson Hamiltonian does not have any 
continuous U(1) symmetry associated with magnon number conservation.

Using the transfer matrix method,~\cite{mackinnon83,ohtsuki04,kramer05} we first
calculated the localization length of a single-particle
eigenstate of a corresponding generalized eigenvalue problem with the randomness.
Due to the bosonic nature, the eigenvalue problem takes a form of
${\mathcal H}_{\rm BdG} |\phi\rangle \equiv \sigma_3 |\phi\rangle E$,
with $\Psi \equiv [\cdots,b_{{\bm j},A},b_{{\bm j},B},\cdots,
b^{\dagger}_{{\bm j},A},b^{\dagger}_{{\bm j},B},\cdots]^T$
and ${\bm H}_{\rm b} \equiv \Psi^{\dagger} {\mathcal H}_{\rm BdG} \Psi$.
$\sigma_3$ in the right hand side is a 2 by 2 diagonal Pauli matrix in the particle-hole space;
$\sigma_3 {\Psi} \equiv [\cdots,+b_{{\bm j},A},+b_{{\bm j},B},\cdots,
-b^{\dagger}_{{\bm j},A},-b^{\dagger}_{{\bm j},B},\cdots]^T$ and $E$ is an
eigenenergy to which $|\phi\rangle$ belongs.
We consider a quasi-one-dimensional (q1d) geometry, where the system is spatially
larger in one direction ($x$-direction) than in the
other ($y$-direction). For every $j_x$ ($j_x= 1,\cdots, L$) with
${\bm j}\equiv (j_x,j_y)$,
the system has a finite width along the $y$-direction; $j_y=1,\cdots,M$
with $M\ll L$.
With $|\phi\rangle  \equiv [|B_{j_x=1}\rangle,\cdots,
|B_{j_x}\rangle,\cdots,|B_{j_x=L}\rangle]^T$ and $|B_{j_x}\rangle \equiv
[\phi_{j_x,j_y=1,A},\cdots,\phi^*_{j_x,j_y=M,B}]^T$, the generalized
eigenvalue equation takes a following $8M \times 8M$ matrix form;
%\begin{eqnarray}
%H_{+} |B_{j_x-1}\rangle + H_{j_x} |B_{j_x}\rangle +
%H_{-} |B_{j_x+1}\rangle = \sigma_3 |B_{j_x}\rangle E.
%\end{eqnarray}
%Note that $H_{j_x}$ differs from one another for different $j_x$
%due to the randomness, while $H_{\pm}$ are same for different $j_x$ with
%$H_{+} \equiv H^{\dagger}_{-}$.
%Equivalently, we have
\begin{widetext}
\begin{eqnarray}
\left(\begin{array}{c}
|B_{j_x+1}\rangle \\
H_{+} |B_{j_x}\rangle \\
\end{array}\right)
= \left(\begin{array}{cc}
- H^{-1}_{-} (H_{j_x}-E\sigma_3) & - H^{-1}_{-} \\
H_{+} & 0 \\
\end{array}\right) \left(\begin{array}{c}
|B_{j_x}\rangle \\
H_{+} |B_{j_x-1}\rangle \\
\end{array}\right) \equiv T_{j_x}
\left(\begin{array}{c}
|B_{j_x}\rangle \\
H_{+} |B_{j_x-1}\rangle \\
\end{array}\right) \equiv M_{j_x}
\left(\begin{array}{c}
|B_{1}\rangle \\
H_{+} |B_{0}\rangle \\
\end{array}\right), \label{A}
\end{eqnarray}
\end{widetext}
and $M_n \equiv \prod^{n}_{j_x=1} T_{j_x}$.
Note that $H_{j_x}$ differs from one another for different $j_x$
due to the on-site randomness, while $H_{\pm}$ are the same for different $j_x$ with
$H_{+} \equiv H^{\dagger}_{-}$. An $8M$ by $8M$ matrix
$T_{j_x}$ has a symplectic feature; $T^{-1}_{j_x} = \tau_y T^{\dagger}_{j_x} \tau_y$,
with $\tau_y$ being a Pauli matrix in the 2-dimensional space subtended
by $|B_{j_x}\rangle$ and $H_{+}|B_{j_x-1}\rangle$. Thus, $T_{j_x}^{\dagger}T_{j_x}$
has a pair of two positive eigenvalues; $\alpha_1,1/\alpha_1, \alpha_2,
1/\alpha_2, \cdots$. The same holds true for $M^{\dagger}_n M_n$.
Call a set of all eigenvalues of an Hermitian matrix
$P_n\equiv M^{\dagger}_nM_n$ as $e^{2n/\lambda_1},e^{-2n/\lambda_1}, e^{2n/\lambda_2},e^{-2n/\lambda_2},\cdots$ with $0<1/\lambda_1<1/\lambda_2<\cdots$.
For sufficiently large $n$, all real $1/\lambda_j$ converge into finite values
(Lyapunov exponents; LE).~\cite{oseledec68} Using the Gram-Schmidt orthonormalization, we
numerically obtained the smallest LE of $P_n$ ($1/\lambda_1$) for larger $n$;
$n = 10^5 \sim 10^6$. $\lambda_1$ is nothing but the largest
localization length of the eigenstate of  
the q1d system at energy $E$.~\cite{mackinnon83} 
We set $E$ inside the topological band gap in the clean limit.
%In fact, there always
%exists an eigenstate of ${\mathcal H}_{\rm BdG}$ which
%has maximum amplitude at $j_x=1$. Eq.~(\ref{A}) dictates that,
%for such an eigenstate, $[|B_{n+1}\rangle,H_{+} |B_{n}\rangle]$ for sufficiently
%large $n$ must be given by a linear combination of $[|B_{1}\rangle,H_{+} |B_{0}\rangle]$
%with its leading coefficient being $e^{-n\lambda_1+i\phi}$. In a word, $1/\lambda_1$ defines
%the largest decay length of the eigenstate along the $x$-direction.

\begin{figure}
\centering
\includegraphics[width=0.48\textwidth]{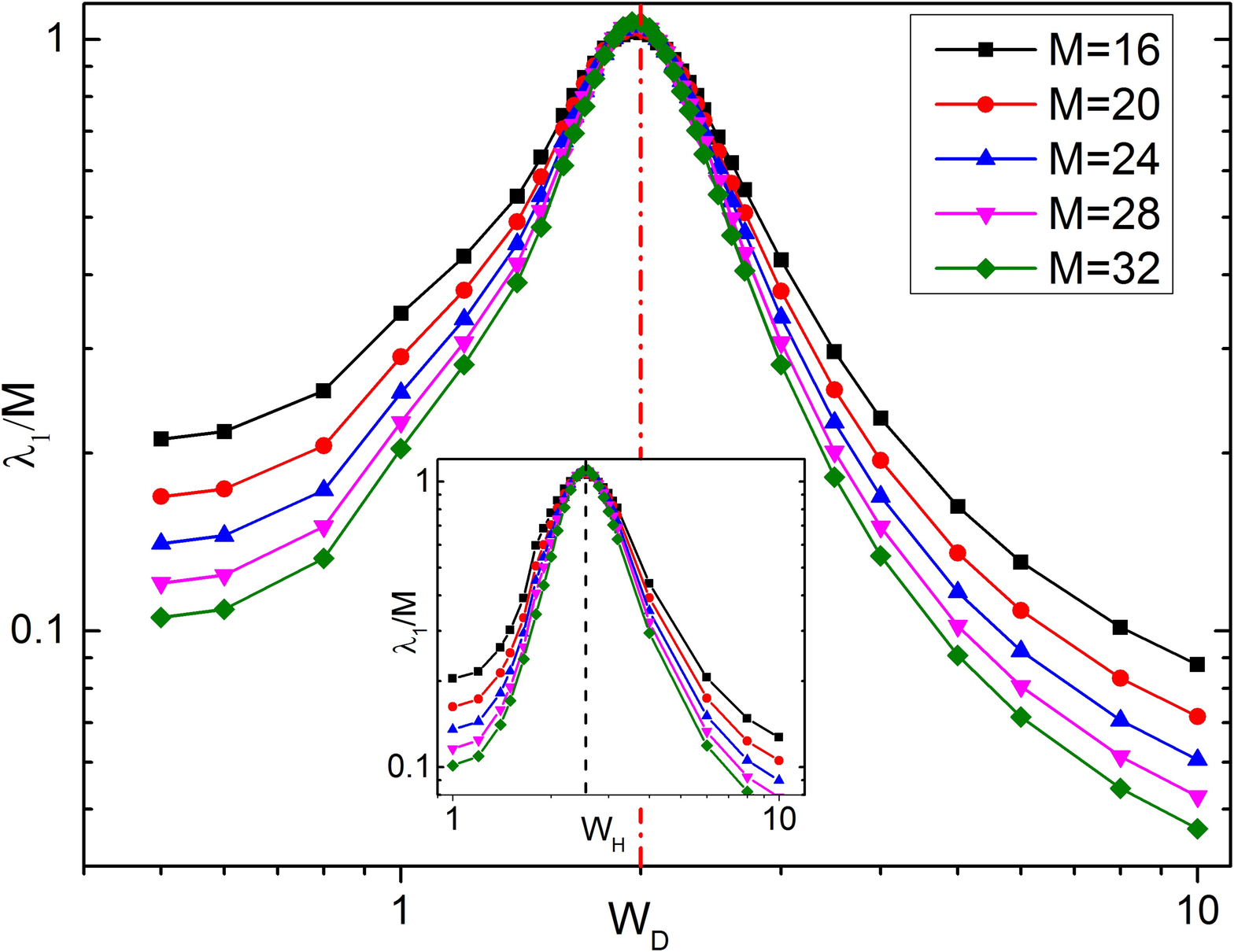}
\caption{(color online) Localization length calculated for different system size $M$
as a function of disorder strength for
the magnetic anisotropy $W_D$, with $H_Z=15$, $DS=2.2$, $JS=1.0$, $J^{\prime}_1S=0.35$, $J^{\prime}_2S=0.28$.
The eigenenergy $E$ is set inside the band gap ($E=3.3$). (Inset)
Localization length as a function of disorder strength for the
Zeeman field $W_H$, with the same set of other parameters. Black broken/red dash 
dotted lines denote 
scale-invariant points of $\lambda_1/M$ in the presence of finite $W_H$/$W_D$
respectively.}
\label{fig1}
\end{figure}

%The smallest LE has been numerically calculated for the various system size $M$,
%where the eigen energy $E$ is
With weaker randomness, the localization length $\lambda_1$ normalized by $M$ decreases
on increasing $M$, suggesting that eigenstates in this regime are all localized due to the
topological band gap (quantum magnon Hall regime).  The same observations
hold true with much stronger randomness, indicating that eigenstates in much
stronger disordered region belong to a conventional Anderson localized regime.
Obtained numerical result (Fig.~\ref{fig1}) shows that these two localized regions are always
separated by a quantum phase transition point where the normalized localization
length barely changes as a function of $M$.
The scale invariant behaviour of $\lambda_1/M$ suggests the existence of a
quantum phase transition similar to an integer quantum Hall plateau-plateau
transition.~\cite{huckestein95,kramer05}

To confirm this, we further calculate the two-terminal magnon conductance $G$ for
the $M \times L$ system
from a transmission matrix ${\bm t} \equiv {\bm T}^{-1}_{11}$
as $hG\equiv {\rm Tr}[{\bm t}^{\dagger}{\bm t}]$.
The transmission matrix is calculated from the transfer matrix;
$[{\bm T}_{11}]_{lm} = {\bm y}^{\dagger}_{l,+} \tau_y \big(\prod^{L}_{j_x=1} T_{j_x}\big) {\bm y}_{m,+}$,
with $\sqrt{|J_{m,+}}{\bm y}_{m,+} \equiv {\bm x}_{m,+}$
and ${\bm J}_{m,+} \equiv {\bm x}^{\dagger}_{m,+} \tau_y {\bm x}_{m,+}$.
%Here
%${\bm x}_{m,+}$ denotes an incident wave defined in the 2-dimensional space subtended by
%$|B_{j_x}\rangle$ and $H_{+}|B_{j_x-1}\rangle$ in eq.~(\ref{A}) with a channel index
%$m$; $m=1,\cdots,4M$. When one choose ${\bm x}_{m,+}$ as eigenstates of eq.~(\ref{A})
%without randomness, the transmission matrix reduces to a unit diagional matrix
%without randomness, justifying the normalization by the flux strength.
Here we choose ${\bm x}_{m,+}$ to be eigenstates of a model
of decoupled one-dimensional chains;
${\bm H}_{\rm lead}=\sum_{\nu=A,B}\sum_{\bm j} (t b^{\dagger}_{(j_x+1,j_y),\nu}
b_{(j_x,j_y),\nu} + {\rm h.c.} + \mu b^{\dagger}_{{\bm j},\nu} b_{{\bm j},\nu})$.
The conductance along the $x$-direction
is calculated both with open ($G_o$) and with periodic
boundary conditions ($G_{p}$) along the $y$-direction.

\begin{figure}[t]
\centering
\includegraphics[width=0.48\textwidth]{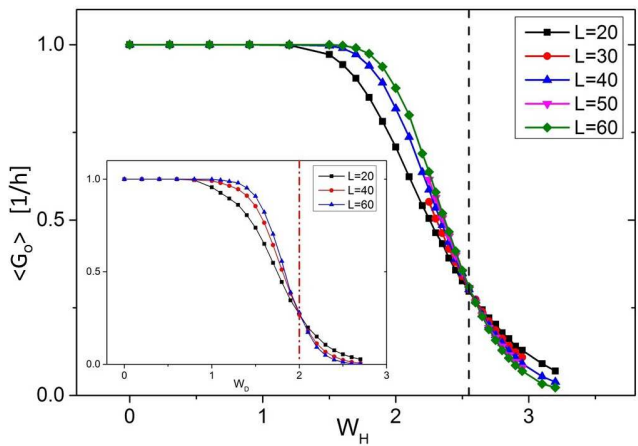}
\caption{(color online) Two-terminal conductance along the $x$-direction
with open boundary condition in the $y$-direction $G_o$ as a function of disorder strength for
the field $W_H$ (inset; as a function of disorder strength for the anisotropy $W_D$).
The lattice geometry is chosen to be a rectangular with $2M=L$ ($L=20$ $\sim$ $60$). Other
parameters are set to be the same as in Fig.~\ref{fig1}. Black broken/red dash dotted lines
denote the scale-invariant points of $\lambda_1/M$ shown in Fig.~\ref{fig1}.}
\label{fig2}
\end{figure}

$G_o$ thus calculated tends to have a finite quantized plateau in the quantum magnon
Hall regime in the thermodynamic limit ($G_o = \frac{1}{h}$),
while showing zero conductance in the conventional
localized regime (Fig.~\ref{fig2}). The quantization in the quantum magnon Hall regime demonstrates a
robust unidirectional magnon transport along the topological chiral edge mode.
The bulk conductance seen by $G_p$ tends to have a finite value only at
the transition point, while zero otherwise in larger system size. These
observations %in combination with the localization length result
lead to the conclusion that the quantum magnon Hall 
regime with the robust chiral edge mode and the conventional
Anderson localized regime without the edge mode are topologically disconnected 
by a direct
transition point with a delocalized bulk state. Importantly, this holds true irrespectively
of the presence of the explicit U(1) symmetry at the Hamiltonian level.

%\begin{figure}[t]
%\centering
%\includegraphics[width=0.48\textwidth]{conductance_H.jpg}
%\caption{(color online) two-terminal conductances as a function of disorder
%strength for the field $W_H$. The  eigen-energy $E=3.3$ and other parameters
%as well as the lattice geometry are set to be same as in Fig.~2.}
%\end{figure}

\begin{figure}[t]
\centering
\includegraphics[width=0.48\textwidth]{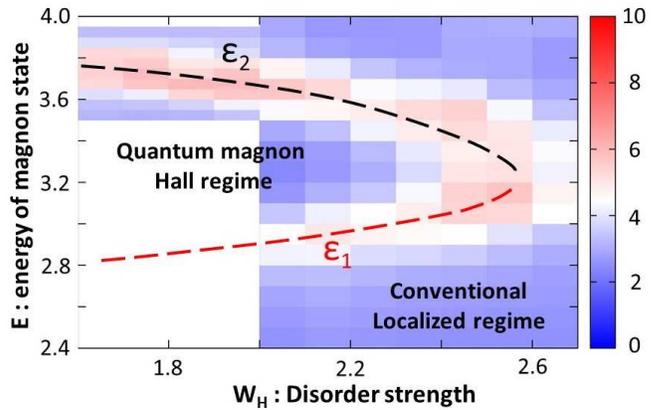}
\caption{(color online) Phase diagram subtended
by the disorder strength $W_H$ and single-magnon energy $E$. The phase boundaries
between quantum magnon Hall regime with the quantized edge conductance ($G_o=\frac{1}{h}$)
and conventional Anderson localized regime, such as 
$E={\mathcal E}_1(W_H)$ and $E={\mathcal E}_2(W_H)$, are
determined from the size dependence of the two-terminal conductance
with periodic boundary condition $G_p$.~\cite{supple}   
Color plot refers to $-\log \sigma$ where
$\sigma$ is a standard deviation of $G_{p}$ with different system size $L$; ${\cal N} \sigma^2
\equiv \sum_{L} (G_{p,L}-\overline{G}_{p})^2$ and ${\cal N}\overline{G}_{p} \equiv \sum_{L} G_{p,L}$ 
with ${\cal N}$ the number of system sizes.
Note that near the scale invariant points, $\sigma^2$ becomes small, hence larger $-{\log} \sigma^2$. 
The parameters are taken to be the same as in Fig.~\ref{fig1}. }
\label{fig3}
\end{figure}

Robustness of chiral magnon edge transports against boson-number-non-conserving
elastic perturbations is a consequence of the energy conservation. Our BdG
type Hamiltonian has a particle-hole symmetry, $\sigma_1 {\mathcal H}_{\rm BdG} \sigma_1
= {\mathcal H}^{*}_{\rm BdG}$, where $\sigma_1$ exchanges particle and hole indices;
$\sigma_1\Psi \equiv [\cdots, b^{\dagger}_{{\bm j},A},b^{\dagger}_{{\bm j},B},\cdots,
b_{{\bm j},A},b_{{\bm j},B}, \cdots]$. Due to this generic symmetry, any
eigenstate $|\phi\rangle$
of ${\mathcal H}_{\rm BdG}$ has its particle-hole counterpart
$|\overline{\phi}\rangle \equiv \sigma_1 (|\phi\rangle)^{*}$. A local perturbation which does not
conserve the boson number can have a finite matrix element between these two, e.g.
$\langle \phi | {\mathcal H}^{\prime} | \overline{\phi}\rangle \ne 0$ with
${\cal H}^{\prime} = b^{\dagger}_{\bm i} b^{\dagger}_{\bm i}  + {\rm h.c.}$.
Physically, however, the hole state and the particle state are different number
states of the same quasi-particle excitation, i.e.
$|\overline{\phi}\rangle \propto |n-1\rangle$ and $|\phi\rangle \propto |n+1\rangle$,
and the scattering process between these two is accompanied by an
energy emission (or absorption) of $2E$, where $E$ is an energy quantum for the
quasi-particle excitation; ${\mathcal H}_{\rm BdG} |\phi\rangle = \sigma_3 |\phi\rangle E$
and ${\mathcal H}_{\rm BdG} |\overline{\phi}\rangle = \sigma_3 |\overline{\phi}\rangle (-E)$.
Thus, any magnon state with $E>0$ cannot be scattered into its hole counterpart
by {\it elastic} scattering. In other words, particle and hole channels are
completely decoupled both in the transmission matrix ${\bm t}$ and in
a reflection matrix ${\bm r}$ in the two-terminal conductance calculation above.
This results in the robustness of the chiral magnon edge
transport even in the presence of boson-number-non-conserving
perturbations. The decoupled nature of particle and hole channels also allows to
define a magnon current even in the absence of the explicit U(1) symmetry in the
magnon Hamiltonian;
%%even though a magnon continuity equation
%%has a source term at the level of an operator equation,
the magnon continuity equation without the source term can be derived from the 
equation of motion for the Green function as far as elastic scattering is concerned.

The robust chiral edge conductance in the quantum magnon Hall regime 
indicates that the Hall regime with the quantized edge conductance 
($G_o=\frac{1}{h}$) is always disconnected from the conventional localized regime
by a direct transition with delocalized bulk states.~\cite{halperin82} 
This is indeed the case with a phase diagram subtended by the 
disorder strength $W_H$ and the single-magnon energy $E$ (Fig.~\ref{fig3}).
For a fixed $W$, the Hall regime is encompassed by
the two direction transition points at $E={\mathcal E}_1(W)$ and
${\mathcal E}_2(W)$. For ${\mathcal E}_1(W)<E<{\mathcal E}_2(W)$,
$G_o$ is quantized and $G_p$ vanishes in the thermodynamic limit. For the other region,
both $G_o$ and $G_p$ tend to vanish in a larger system size.~\cite{supple}

\begin{figure}[t]
\centering
\includegraphics[width=0.48\textwidth]{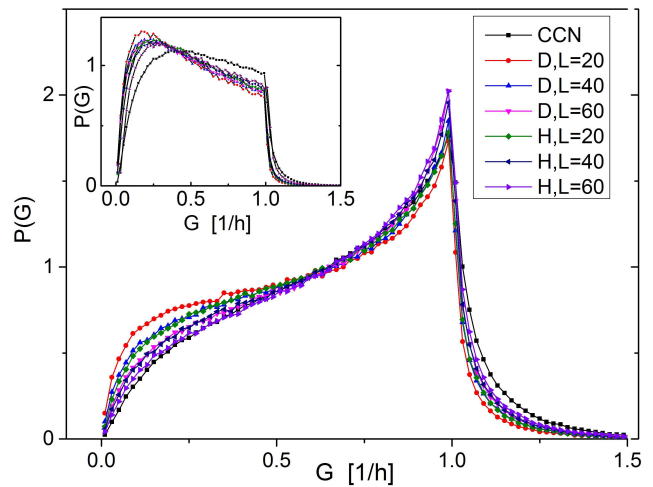}
\caption{(color online) Critical conductance distributions of $G_o$ with $M=L$
with different system size ($L=20,40,60$) and different types of randomness
(either $W_H$ or $W_D$). The black square points show the critical conductance
distribution of the Chalker-Coddington network (CCN) model. 
The energy $E$ and the other parameters are the same as
in Figs. 1 and 2, respectively. The critical values of $W_H$ and $W_D$ are chosen
to be a scale invariant point of $G_o$. (Inset) Critical conductance distribution of $G_p$ and its
comparison with that of the CCN model.}
\label{fig4}
\end{figure}

A finite-size scaling analyses of $G_o$ (of Fig.~\ref{fig2}) near the transition point ($W=W_c$) is
carried out based on $G_{o} (L^{1/\nu}(W-W_c),L^{-|y|}) \simeq
 G_c + \sum^3_{n=1} c_n (W-W_c)^n L^{n/\nu}
+ b L^{-|y|}$, with $\nu$ the critical exponent, $y$ a scaling dimension of a leading-order
irrelevant scaling field at the critical point and $c_n$ fitting parameters.~\cite{slevin14}
%Different fitting schemes
%give $\nu=2.44 \pm 0.08$ (for $L=20,30,40,50,60$ without $c_4$;
%the number of data point $N =66$, $\chi^2=102$), %, Goodness $=0.003$),
%$\nu=2.42 \pm 0.02$ (with $L=30,40,50,60$ without $y$; $N=53$, $\chi^2=50$) %, Goodness $=0.61$)
%, $\nu=2.70 \pm 0.08$ (for $L=30,40,50,60$ without
%$c_4$; $N=51$, $\chi^2=36$) %, Goodness $=0.933$).
%The first two values are closed to
%known values of the exponent of the two-dimensional integer quantum Hall plateau-plateau
%transition, while the latter is to that of the two-dimensional symplectic class.
For $L=20,30,40,50,60$, the 95\% confidence interval of $\nu$ is [2.28,2.60] with goodness of fit $Q=0.003$.
By omitting the smallest size, % i.e., for $L=30,40,50,60$,
the estimate is [2.54, 2.86] with goodness of fit $Q=0.93$.
Though being consistent with the recent estimate of $\nu$ of the quantum Hall university
class ($\nu\approx 2.59$),~\cite{slevin09,obuse10,amado11,fulga11,dahlhaus11,obuse12} 
the error bars are too large to conclude this affirmatively. To this end, we further
calculated distributions of the conductances at the critical point (Fig.~\ref{fig4}).
The distributions have striking similarities to the critical
conductance distributions of the two-dimensional Chalker-Coddington network model,~\cite{chalker88} 
which strongly suggests that the direct transition belongs to the quantum Hall universality class.

Based on these knowledge, let us finally characterize an edge-mode contribution to
thermal magnon Hall conductivity $\kappa_{xy}$ in the disordered quantum magnon Hall regime.
To this end, we impose an open/periodic boundary condition along the
$y$/$x$-direction, introduce a temperature gradient along the
$y$-direction, and calculate an energy current along the
$x$-direction. The energy Hall current is given as a function of the disorder strength $W$.
For a given $W$, the system has sub-extensive number of chiral edge modes within
${\mathcal E}_1(W)< E<{\mathcal E}_2(W)$, where $G_o=\frac{1}{h}$. % in ${\mathcal E}_1<E<{\mathcal E}_2$,
Thus, the edge modes within $[E,E+dE]$ give a magnon Hall current density of $\frac{dE}{h M}$,
with $M$ the system size along the $y$-direction.
The energy Hall current density due to these chiral edge modes around $y=M$ with
higher temperature $T_H$ and that around $y=0$ with lower temperature $T_L$ are therefore,
\begin{align}
I^{x,E}_{H/L} &= \pm \frac{1}{h M} \int^{{\mathcal E}_2(W)}_{{\mathcal E}_1(W)} g(E,T_{H/L}) E dE, \nn
%I_{L} &= - \frac{1}{\hbar} \int^{E_2(W)}_{E_1(W)} g(E,T_{L}) E dE, \nn
\end{align}
respectively with Bose function $g(E,T) \equiv 1/[e^{E/k_B T}-1]$.
A sum of these two is proportional to the temperature gradient
$\Delta_y T \equiv T_H-T_L$;
$I^{x,E}_t \equiv I^{x,E}_H + I^{x,E}_L = \kappa^{\rm edge}_{xy} \partial_y T
= \kappa^{\rm edge}_{xy} \Delta_y T/M$.
The thermal Hall conductivity takes a form;
\begin{align}
&\kappa^{\rm edge}_{xy}(W) = - \frac{k^2_B T}{h} \Big( C_2\big(g({\mathcal E}_2,T)\big)
-C_2\big(g({\mathcal E}_1,T)\big)\Big), \label{kappaxy}
\end{align}
with $T \equiv \frac{T_H+T_L}{2}$ and $C_2(x)$ is a non-analytic function; 
$C_2(x) \equiv \int^{x}_{0} \big(\ln \frac{1+t}{t}\big)^2 dt$. 
The above argument can be easily generalized into
generic quantum magnon Hall systems with disorders.~\cite{supple} Note also that the edge-mode
contribution dominates total $\kappa_{xy}$ in a system with the quasi-one-dimensional
geometry ($M\ll L$), where a bulk contribution diminishes as $e^{-aL/M}$ ($a$ being a constant
of the order of $1$) due to the localization effect in one-dimensional systems.

In this rapid communication, we studied low-energy magnon bands in a two-dimensional
spin ice model with disorders. We show that the magnon bands with
disorders undergo a direct transition from an integer quantum magnon Hall regime to
a conventional magnon localized regime. The critical conductance distributions at
the transition point suggest that the direct transition belongs to quantum Hall universality
class. The obtained result can be tested by standard 
microwave antennas experiments.~\cite{supple} 
Based on the edge magnon transport picture, we give a generic expression for
thermal magnon Hall conductivity in {\it disordered} quantum magnon Hall systems.
The obtained expression is qualitatively consistent with an expression of
thermal magnon Hall conductivity in the clean limit, previously obtained based on
the linear response theory.~\cite{matsumoto11a,matsumoto11b,matsumoto14,qin11,qin12,supple} 

The authors thank Junren Shi for fruitful discussions. This work was supported by JSPS KAKENHI
Grants No. 15H03700 and No. 24000013 and by NBRP of China (Grant No. 2015CB921104).

\newpage

\widetext

\section{\textbf{Supplemental Materials for ``Integer Quantum Magnon Hall Plateau-Plateau Transition in
a Spin Ice Model''}}

\section{two-dimensional spin ice model under strong out-of-plane field}
\begin{figure}[t]
\centering
\includegraphics[width=0.40\textwidth]{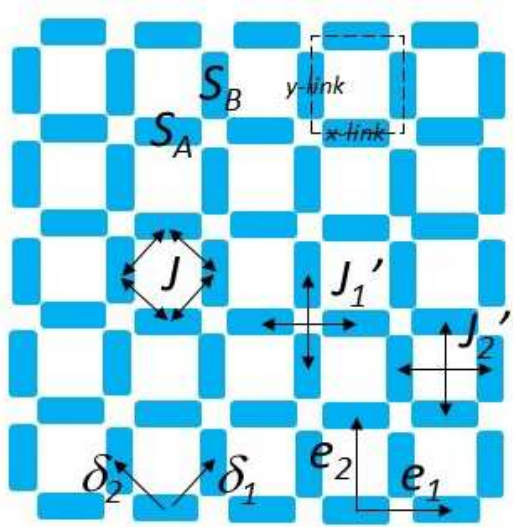}
\caption{(color online) Square-lattice spin ice model with two inequivalent ferromagnetic spins;
${\bm S}_{{\bm i}\in A}$ and ${\bm S}_{{\bm i}\in B}$ are on the center of the $x$-link and $y$-link
of the square lattice respectively. The nearest neighbor dipole coupling strength is denoted by $J$,
the next-nearest neighbor dipole coupling strengths are denoted by $J^{\prime}_{1}$ and
$J^{\prime}_2 (<J^{\prime}_1)$. $e_1$ and $e_2$ are the primitive lattice vectors of the
square lattice. $\delta_{1}$ and $\delta_2$ connect the nearest neighbor spins; $\delta_1\equiv \frac{e_1+e_2}{2}$
and $\delta_2\equiv \frac{e_1-e_2}{2}$.}
\label{s-fig1}
\end{figure}

A magnetic system considered consists of two inequivalent ferromagnetic `islands' of the order
of 100 nm size~\cite{s-wang06}; one is centered on a $x$-link of a two-dimensional square lattice and the other is
on a $y$-link (Fig.~\ref{s-fig1}). The ferromagnetic island on the $x$/$y$-link is spatially elongated along the
$x$/$y$-direction respectively. Thus, their magnetic moments prefer to point along the
$\pm x/y$-direction respectively due to the magnetic shape anisotropy.
We model these two ferromagnetic islands as two inequivalent spins on the
$x/y$-link with large magnetic moment $S$ (${\bm S}_{i\in A}$/${\bm S}_{i\in B}$ respectively).
The shape anisotropy is included as an effective single-ion spin-anisotropy energy
such as $-D (S^x_{i\in A})^2$ and $-D (S^y_{i\in B})^2$.
Ferromagnetic moments are coupled with one another via the magnetic dipole-dipole interaction~\cite{s-wang06},
so do spins in the spin model. Based on the Holstein-Primakoff mapping, the corresponding quadratic
magnon Hamiltonian is derived as in Eq.~(2) (in the main text).
For simplicity, we include only dipole couplings between the nearest neighbor spins
(denoted by $J$: see Fig.~\ref{s-fig1}) and between next nearest neighbor spins
(denoted by $J^{\prime}_{A,1}=J^{\prime}_{B,2}=J^{\prime}_1$ or
$J^{\prime}_{A,2}=J^{\prime}_{B,1}=J^{\prime}_2$: see Fig.~\ref{s-fig1}).
The ratio among these three are chosen to be consistent
with the $1/r^3$ dipolar coupling strength (see below).

Without randomness (${\bm d}_j\equiv {\bm h}_{j}\equiv 0$), the quadratic Hamiltonian given in Eq.~(2) in the main text can be
Fourier-transformed, %given in the momentum space as follows,
\begin{align}
{\bm H}_b &= \frac{1}{2}\sum_{\bm k} \left(\begin{array}{cccc}
b^{\dagger}_{{\bm k},A} & b^{\dagger}_{{\bm k},B} & b_{{-{\bm k}},A} & b_{-{\bm k},B} \\
\end{array}\right) \cdot {\bm H}_{\rm BdG}({\bm k}) \cdot
\left(\begin{array}{c}
b_{{\bm k},A} \\
b_{{\bm k},B} \\
b^{\dagger}_{{-{\bm k}},A} \\
b^{\dagger}_{-{\bm k},B} \\
\end{array}\right),   \nn \\
{\bm H}_{\rm BdG}({\bm k}) &= a_0({\bm k}) {\bm \gamma}_0 + a_2({\bm k}) {\bm \gamma}_2
+ a_{14}({\bm k}) {\bm \gamma}_{14} + a_{23}({\bm k}) {\bm \gamma}_{23} +
a_{45}({\bm k}) {\bm \gamma}_{45}, \label{bdg-k}
\end{align}
with
\begin{align}
{\bm \gamma}_{j} \equiv {\bm \sigma}_{j} \otimes {\bm \tau}_1, \
{\bm \gamma}_4 \equiv {\bm \sigma}_0 \otimes {\bm \tau}_2, \
{\bm \gamma}_5 \equiv {\bm \sigma}_0 \otimes {\bm \tau}_3, \
{\bm \gamma}_{\mu\nu} \equiv i {\bm \gamma}_{\mu}{\bm \gamma}_{\nu},
\end{align}
where $j=1,2,3$ and $\mu,\nu=1,\cdots,5$. 2 by 2 Pauli matrix ${\bm \sigma}_{j}$
is for the particle-hole space, while 2 by 2 Pauli matrix ${\bm \tau}_{j}$
is for the A and B sublattices. Coefficients in Eq.~(\ref{bdg-k}) are
\begin{align}
a_{0}({\bm k}) &= H_Z - DS - 4JS - J^{\prime} S (\cos k_x + \cos k_y) - 4J^{\prime} S, \label{eq1} \\
a_{2}({\bm k}) &= -6JS \sin \frac{k_x}{2} \sin \frac{k_y}{2},  \ \ a_{14}({\bm k})  = DS, \label{eq2} \\
a_{23}({\bm k}) & = 3 J^{\prime} S (\cos k_x - \cos k_y),  \ \
a_{45}({\bm k})  = 2JS \cos \frac{k_x}{2} \cos \frac{k_y}{2}, \label{eq3}
\end{align}
where we put $J^{\prime}_1=J^{\prime}_2 \equiv J^{\prime}$
for simplicity. The 4 by 4 matrix is diagonalized at every ${\bm k}$
by a paraunitary transformation~\cite{s-colpa78}
\begin{eqnarray}
{\bm T}({\bm k})
= e^{-i\theta_1 {\bm \gamma}_3} e^{i\frac{\pi}{4} {\bm \gamma}_4} e^{\mu_1 {\bm \gamma}_1}
e^{-i\theta_2 {\bm \gamma}_{3}} \left(\begin{array}{cccc}
\cosh \mu_2 & 0 & -i \sinh \mu_2 & 0 \\
0 & \cosh \mu_3 & 0 & - i\sinh \mu_3 \\
i\sinh \mu_2 & 0 & \cosh \mu_2 & 0 \\
0 & i\sinh \mu_3 & 0 & \cosh \mu_3 \\
\end{array}\right),
\end{eqnarray}
where $\theta_1$, $\theta_2$, $\mu_1$, $\mu_2$ and $\mu_3$ are defined by
$a_0$, $a_{2}$, $a_{14}$, $a_{23}$ and $a_{45}$ as follows;
\begin{align}
\sin 2\theta_1 &= \frac{a_{23}}{\sqrt{a^2_{23} + a^2_2}}, \ \  \cos 2\theta_1 = \frac{a_2}{\sqrt{a^2_{23} + a^2_2}},
\label{eq4} \\
\sin 2\theta_2 &= \frac{\sqrt{a^2_{23}+a^2_2}\frac{a_{14}}{\sqrt{a^2_0 - a^2_{14}}}}{\sqrt{a^2_{45} + \Big(\sqrt{a^2_{23}+a^2_{2}}\frac{a_{14}}{\sqrt{a^2_0 - a^2_{14}}}\Big)^2 }}, \ \ \cos 2\theta_2 =\frac{a_{45}}{\sqrt{a^2_{45} + \Big(\sqrt{a^2_{23}+a^2_{2}}\frac{a_{14}}{\sqrt{a^2_0 - a^2_{14}}}\Big)^2 }}
, \label{eq5} \\
\sinh 2 \mu_1 & = \frac{a_{14}}{\sqrt{a^2_0 - a^2_{14}}}, \ \ \cosh 2\mu_1 = \frac{a_{0}}{\sqrt{a^2_0 - a^2_{14}}},
\label{eq6} \\
\sinh 2\mu_2 &= \frac{b_{24}}{\sqrt{(b_0+b_5)^2-b^2_{24}}}, \ \
\cosh 2\mu_2 = \frac{b_{0}+b_5}{\sqrt{(b_0+b_5)^2-b^2_{24}}}, \label{eq7} \\
\sinh 2\mu_3 &= \frac{-b_{24}}{\sqrt{(b_0-b_5)^2-b^2_{24}}}, \ \
\cosh 2\mu_3 = \frac{b_{0}-b_5}{\sqrt{(b_0-b_5)^2-b^2_{24}}}, \label{eq8}
\end{align}
with
\begin{eqnarray}
b_0 \equiv \sqrt{a^2_0 - a^2_{14}}, \
b_{24} \equiv \sqrt{a^2_{23} + a^2_2} \frac{a_{0}}{\sqrt{a^2_0 - a^2_{14}}}, \
b_{5} \equiv \sqrt{a^2_{45} + \Big(\sqrt{a^2_{23} + a^2_2} \frac{a_{14}}{\sqrt{a^2_0 - a^2_{14}}}\Big)^2}. \label{eq9}
\end{eqnarray}
In terms of the transformation, the BdG Hamiltonian is paraunitary equivalent to a diagonal matrix,
\begin{eqnarray}
{\bm T}^{\dagger}({\bm k})\cdot {\bm H}_{\rm BdG}({\bm k}) \cdot {\bm T}({\bm k}) = \left(\begin{array}{cccc}
E_{+}({\bm k}) &  & & \\
& E_{-}({\bm k}) & &  \\
& & E_{+}(-{\bm k}) & \\
& & & E_{-}(-{\bm k}) \\
\end{array}\right), \ \  {\bm T}^{\dagger}({\bm k}) \cdot {\bm \sigma}_3 \otimes {\bm \tau}_0
\cdot {\bm T}({\bm k}) =  {\bm \sigma}_3 \otimes {\bm \tau}_0, \nn
\end{eqnarray}
where the two magnon energy bands are even functions in ${\bm k}$;
\begin{eqnarray}
E_{\pm} ({\bm k}) \equiv \sqrt{\big(b_0({\bm k}) \pm b_5({\bm k})\big)^2 - b^2_{24}({\bm k})}. \label{energy}
\end{eqnarray}
Note that $H_Z$ is sufficiently large that the lower magnon band $E_{-}({\bm k})$ is fully gapped;
$E_{-}({\bm k})>0$ for $\forall {\bm k}$. This also allows that
$a^2_{0}-a^2_{14}>0$ and $b_0 > b_5$.  In the absence of the next-nearest neighbor
dipolar interaction $J^{\prime}=0$ ($a_{23}=0$), two bands form a band touching with a massless
Dirac dispersion at ${\bm k}=(0,\pi)$ and $(\pi,0)$, where $a_{45}=a_{2}=b_5=0$. The Dirac dispersions
acquire a finite mass by non-zero next-nearest neighbor dipolar interaction and the sign of the mass
is determined by that of $J^{\prime}$. Due to this mass acquaintance, the upper magnon band ($E_{+}({\bm k})$)
and  lower magnon band ($E_{-}({\bm k})$) have $\pm 1$ Chern integer respectively
for $J^{\prime} > 0$. When $J^{\prime}$ changes its sign from positive to negative,
these two integers change into $\mp 1$ respectively.

\begin{figure}[t]
\centering
\includegraphics[width=0.90\textwidth]{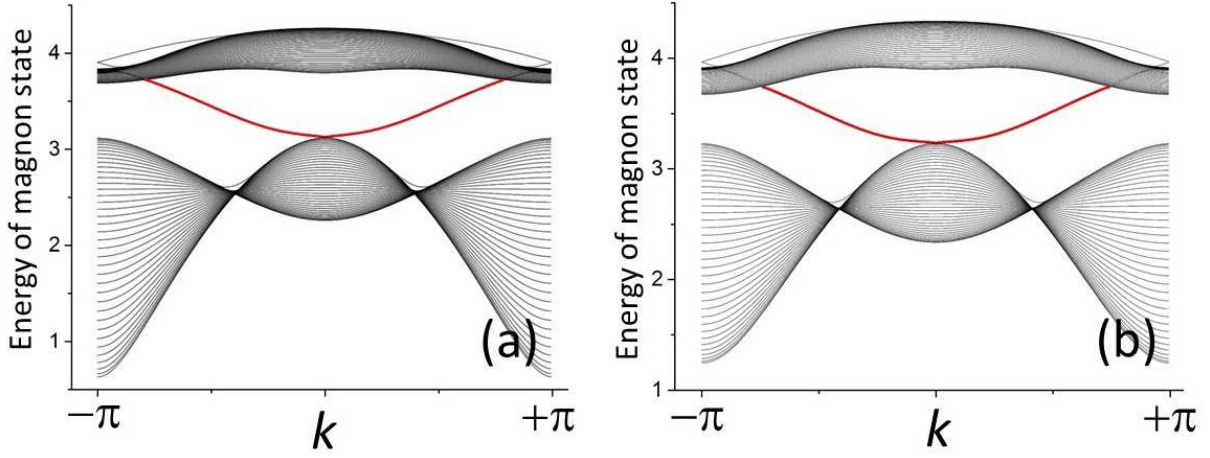}
\caption{(color online) Energy band dispersions of magnon states as function of the
wave vector $k$, calculated with open/periodic boundary condition along the $y$/$x$-direction.
Black colored dispersions are for bulk modes, and red colored dispersions are
for the chiral edge modes; those at $k<0$ are localized at $y=M (>0)$ and those at $k>0$ are localized
at $y=0$. (a) $J^{\prime}_1S=J^{\prime}_{2}S=0.35$, $H_Z=15.8$, $DS=2.2$, $JS=1.0$,
(b) $J^{\prime}_1S=0.35, J^{\prime}_{2}S=0.28$, $H_Z=15.8$, $DS=2.2$, $JS=1.0$.}
\label{s-fig2}
\end{figure}

To calculate the Chern integer for the upper magnon band 
directly~\cite{s-thouless82,s-kohmoto85,s-shindou13a,s-engelhardt15,s-furukawa15,s-xu16},
look into the first column of the paraunitary matrix ${\bm T}({\bm k})$, which is nothing but a periodic part
 of the Bloch wavefunction for the upper magnon band,
\begin{eqnarray}
{\bm T}({\bm k}) = \left(\begin{array}{cccc}
\frac{u_{1,+}({\bm k})}{N_+} &\cdots & \cdots & \cdots \\
\frac{u_{2,+}({\bm k})}{N_+} & \cdots & \cdots &\cdots  \\
\frac{v_{1,+}({\bm k})}{N_+} & \cdots & \cdots & \cdots \\
\frac{v_{2,+}({\bm k})}{N_+} & \cdots & \cdots & \cdots \\
\end{array}\right). \label{T}
\end{eqnarray}
where $u_{1,+}$ and $u_{2,+}$ ($v_{1,+}$ and $v_{2,+}$) are connected with each other
by the $C_4$ rotation ($(k_x,k_y) \rightarrow
(-k_y,k_x)$, exchanges A and B sublattices). $u_{j,+}$ and $v_{j,+}$ are connected with each other
by a particle-hole transformation, which is generic in any quadratic boson Hamiltonian;
${\bm \sigma}_1 \cdot {\bm H}_{\rm BdG}({\bm k}) \cdot
{\bm \sigma}_1 = {\bm H}^{*}_{\rm BdG}(-{\bm k})$. $u_{1,+}({\bm k})$ is given by
\begin{align}
u_{1,+}({\bm k}) &= \cos (\theta_1 + \theta_2)
\cosh \mu_1 \cosh \mu_2 + \sin (\theta_1 - \theta_2)
\sinh \mu_1 \sinh \mu_2 \nn \\
& \hspace{2cm} + i \big( \sin (\theta_1 - \theta_2)
\cosh \mu_1 \cosh \mu_2 + \cos (\theta_1 + \theta_2)
\sinh \mu_1 \sinh \mu_2\big). \label{u1}
\end{align}
while the other three are obtained from this by the $C_4$ rotation or by the
particle-hole transformation. ${\bm T}({\bm k})$ is given by a proper normalization; $N^2_+ \equiv
u^2_{1,+} + u^2_{2,+} -v^2_{1,+} - v^2_{2,+}$. By using Eqs~.(\ref{eq1}-\ref{eq9}), one can see that
$u_{1,+}$ (and also $v_{1,+}$) has a zero only at ${\bm k}=(0,\pi)$, while $u_{2,+}$ and $v_{2,+}$ have a zero at
${\bm k}=(\pi,0)$. Thus, we expand $u_{1,+}({\bm k})$ with respect to small
${\bm q}$ with ${\bm k}\equiv (0,\pi)+{\bm q}$;
\begin{eqnarray}
u_{1,+}((0,\pi)+{\bm q}) = {\bm X}\cdot {\bm q} + i {\bm Y}\cdot {\bm q} + {\cal O}(q^2),
\end{eqnarray}
where ${\bm X}\equiv (-a,-b)$, ${\bm Y} \equiv (a,-b)$. $a$ and $b$ are calculated as follows,
\begin{align}
a &= \frac{d\theta_1}{d k_x}|_{{\bm k}=(0,\pi)} \!\
\Big(\cosh \mu_1 \cosh \mu_2
- \sinh \mu_1 \sinh \mu_2\Big) =  \frac{J}{4J^{\prime}} \!\ \Big(\cosh \mu_1 \cosh \mu_2
- \sinh \mu_1 \sinh \mu_2\Big) > 0 \nn \\
b &= \frac{d\theta_2}{d k_y}|_{{\bm k}=(0,\pi)} \!\
\Big(\cosh \mu_1 \cosh \mu_2
+ \sinh \mu_1 \sinh \mu_2\Big) = \frac{JS \sqrt{a^2_0-a^2_{14}}}{3\sqrt{2}J^{\prime}S DS}
\Big(\cosh \mu_1 \cosh \mu_2
+ \sinh \mu_1 \sinh \mu_2\Big)> 0. \nn
\end{align}
for $J^{\prime}>0$ and $D>0$.
Note that ${\bm X}\times{\bm Y}=2ab>0$. Thus, a phase of $u_{1,+}({\bm k})$, i.e. $\theta_{1,+}({\bm k})$
with $u_{1,+}\equiv e^{i\theta_{1,+}}|u_{1,+}|$, acquires $+2\pi$ phase holonomy,
whenever ${\bm k}$ rotates once around $(0,\pi)$ anti-clockwise. This dictates that
the Chern integer for the upper band is $+1$ (that for the lower band is $-1$
due to the sum rule~\cite{s-shindou13a}).
The non-zero topological integers for these two magnon bands result
in a topological chiral magnon edge mode within the band
gap~\cite{s-halperin82,s-hatsugai93,s-shindou13a,s-engelhardt15,s-furukawa15,s-xu16}.
The sign of the integer dictates that the mode has a chiral
dispersion with the anti-clockwise rotation when
viewed from $+z$ direction and the out-of-field is along $+z$ direction (``right-handed''
chiral mode).
So far, we assume that $J^{\prime}_1=J^{\prime}_2=J^{\prime}$. For
$J^{\prime}_2=0.8 J^{\prime}_1$, we confirmed numerically that the same
band gap with the chiral edge mode persists (Fig.~\ref{s-fig2}).

\section{A phase boundary between the quantum magnon Hall regime and
conventional magnon localized regime}

\begin{figure}[t]
\centering
\includegraphics[width=0.95\textwidth]{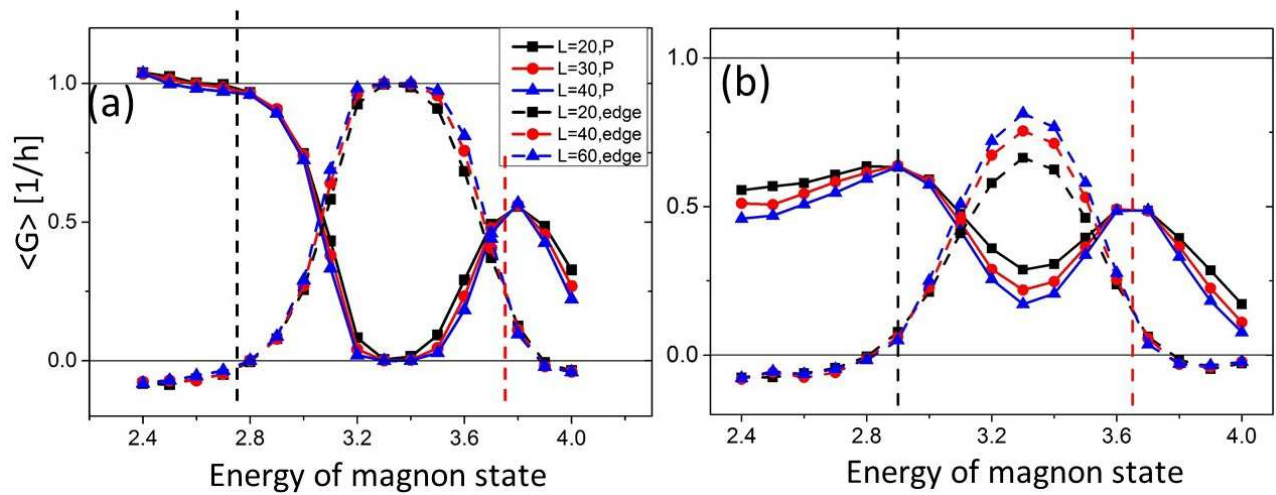}
\caption{(color online) Two-terminal conductance as a function of energy of magnon state $E$ for
several $W$ ((a) $W=1.3$ and (b) $W=2.1$), calculated for the $L\times L$ system.
Solid lines with different colors denote the conductance along the $x$-direction with
periodic boundary condition along the $y$-direction $G_p$; $L=20$ (black), $L=30$ (red), $L=40$ (blue).
Broken lines with different colors denote $G_o-G_p$, where $G_o$ is the conductance with open
boundary condition along the $y$-direction. The other parameters are the same as those given
in the caption of Fig. 1 and 4 in the main text. Two direct transition points are identified
as a scale-invariant point of $G_p$; a red colored dotted line for $E={\cal E}_2(W)$ and black colored
dotted line for $E={\cal E}_1(W)$. Note that, for ${\cal E}_1(W)<E<{\cal E}_2(W)$, $G_o-G_p$ has a
tendency to take the quantized value ($1/h$) in the thermodynamic limit. }
\label{s-fig3}
\end{figure}

A phase diagram in the main text (Fig.~3) has the quantum magnon
Hall regime and conventional magnon localized regime. The boundary between
these two regions is identified as a scale-invariant point of the two-terminal
conductance calculated with the periodic boundary condition; $G_p$. Fig.~\ref{s-fig3}
shows $G_p$ as a function of the single-particle
(magnon) energy $E$ for several $W$. Thereby, we found two such scale-invariant
points (one at $E={\cal E}_1(W)$ specified by a black dotted line in Fig.~\ref{s-fig3}
and the other at $E={\cal E}_2(W)$ by a red dotted line). For ${\cal E}_1(W)<E<{\cal E}_{2}(W)$
an ``edge conductance'' characterized by $G_o-G_p$ has
a tendency to take the quantized value ($1/h$) in the thermodynamic limit
($G_o$ is the conductance along the $x$-direction
with the open boundary condition along the $y$-direction) .
For $E<{\cal E}_1(W)$ or $E>{\cal E}_2(W)$, the edge conductance goes to zero.
From these observations, we regard the former region as the quantum magnon Hall regime
and the latter as the conventional magnon localized regime.

\section{microwave antennas experiment}

\begin{figure}
\centering
\includegraphics[width=0.85\textwidth]{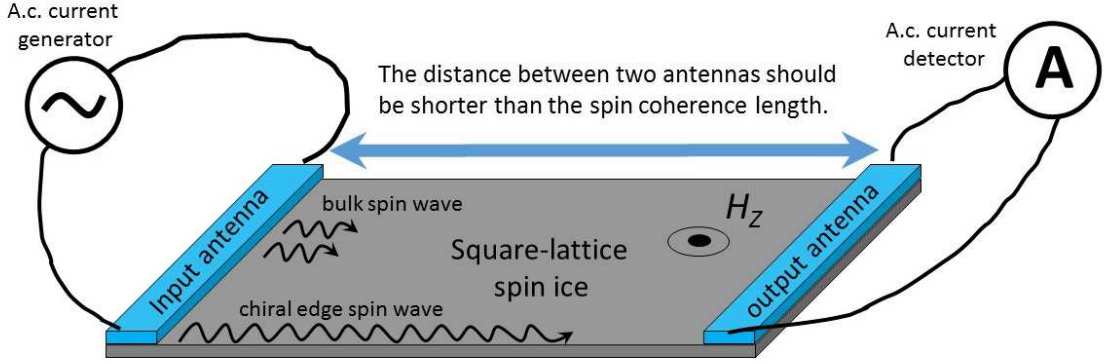}
\caption{A schematic picture for microwave antennas experiment. Two blue-colored
bars denote the coplanar waveguide. The gray-colored plate denotes the two-dimensional
patterned ferromagnetic film (square-lattice spin ice).}
\label{s-fig4}
\end{figure}

The two terminal magnon conductance calculated in the main text can be measured in a
standard microwave experiment commonly used for spin wave experiments~\cite{s-serga10}.
The experiment consists of two microstrip microwave antennas attached to
the two-dimensional square-lattice spin ice system (Fig.~\ref{s-fig4}).
The two antennas are spatially separated from each other shorter than a spin coherent length,
over which spin wave propagates without an energy dissipation. Note that the spin
coherent length in ferromagnetic insulator such as YIG can be over millimeters,
while it is at most on the order of several micrometer in ferromagnetic metals.

The role of the first antenna is for spin wave excitation and
that of the second antenna is for its detection. An a.c. electric current with a
frequency in the microwave regime (let us call this as `external frequency') is introduced
in the first antenna (`input signal'). The current locally excites spin wave with the same
external frequency. The spin wave propagates through the magnonic crystal system, and,
after a certain time, the spin wave reaches the second antenna, where an a.c. electric
current is induced (`output signal').

The two terminal magnon conductance studied in the main text
corresponds to a transmission ratio between the
output electric current and input current. The ratio can be
obtained as function of the external frequency within the microwave regime~\cite{s-serga10}.
When the frequency and the disorder
strength are chosen inside the quantum magnon Hall
regime, the transmission ratio is finite (see Fig.~3 of the main text; the
frequency corresponds to  energy of magnon states in the figure).
Especially, the ratio is dominated by chiral edge spin wave transport,
when the distance between two antennas is longer than the localization length.
Namely, the bulk spin wave excited by the first antenna dies off quickly before it
reaches the second antenna due to its finite localization length. Meanwhile,
the chiral edge spin wave
excited by the first antenna travels along the edge without being backward scattered.

When the frequency and the disorder are in the conventional magnon localized regime,
the transmission ratio reduces dramatically. The
ratio becomes exponentially small, if the localization length is much
shorter than the distance between the two antennas. Accordingly, the quantum phase
transition from the quantum magnon Hall regime to conventional magnon localized
regime can be experimentally measurable through the dramatic reduction of the
transmission ratio as a function of either the external frequency or the disorder
strength.

Note that the distance between the two antennas must be shorter
than a finite spin coherence length (Fig.~\ref{s-fig4}).
The finite distance between the two antennas
may result in a blurred change of the transmission ratio at
the phase transition point. Nonetheless, the spin ice model made out of
ferromagnetic insulator such as YIG allows a very large distance between the two
antennas, e.g. 8mm in YIG~\cite{s-serga10}. Since a typical localization length would be at largest on the
order of micrometer scale~\cite{s-evers15}, the very large spin coherence 
length in YIG may even enable us to study the critical properties of the 
quantum phase transition.

\section{thermal magnon Hall conductivity in generic disordered quantum magnon Hall systems and
its relation to the thermal magnon Hall conductivity in the clean limit}
\begin{figure}
\centering
\includegraphics[width=0.85\textwidth]{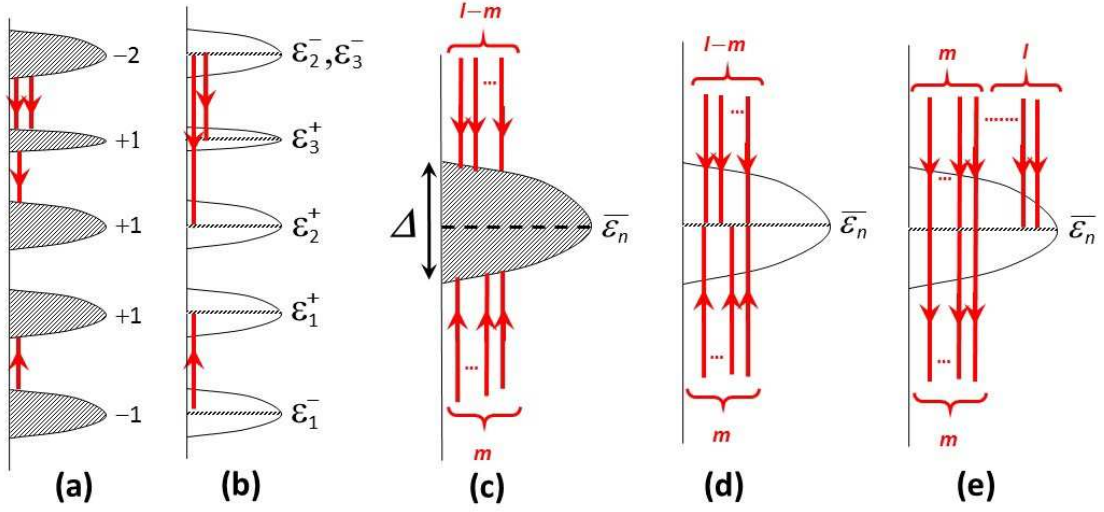}
\caption{Schematic picture for bulk magnon bands and chiral edge modes in the
clean limit (a,c) and with disorders (b,d,e,f). The Chern integers for each magnon band
are shown in (a) such as $-1,+1,+1,+1,-2$ (from below). A right/left-handed chiral edge
mode is depicted by a red line with up/down-headed arrow respectively.
Gray shadow regions denote extended bulk states, while the white region 
in the bands corresponds to mobility gaps. (c,d) the $n$-th band with
${\rm Ch}(n) = l$, $\sum^{n-1}_{j=1}{\rm Ch}(j)=-m$ and $l>m>0$,
(e) When the $n$-th band with ${\rm Ch}(n) = l$, $\sum^{n-1}_{j=1}{\rm Ch}(j)=m$ and 
$l>0$, $m>0$ is disordered.}
\label{s-fig5}
\end{figure}

In the main text, we have studied only the model with two magnon bands.
A realistic material may have more than two magnon bands, which have non-zero
quantized Chern integers. Our study
as well as established knowledge on interplays between localization effect and
quantum Hall physics~\cite{s-prange} suggests that even small disorder makes all
these bulk magnon bands localized except for delocalized bulk states at
respective band center (Fig.~\ref{s-fig5}(a,b)). A pair of two delocalized bulk states
bound a mobility gap, inside which a topological chiral edge mode lives (Fig.~\ref{s-fig5}(b)).
For the two-band model studied in the main text,
the bulk delocalized states at $E={\cal E}_{2}$ and $E={\cal E}_1$
encompass the mobility gap, inside which the chiral edge mode lives.
As in Fig.~3 of the main text, the edge mode disappears when a pair of the
two delocalized bulk states fall into the same energy.

For the generic situation described above, we can employ the same argument as
in the main text, to derive an edge contribution to the thermal Hall conductivity,
\begin{eqnarray}
\kappa_{xy} = - \frac{k^2_B T}{h} \sum_{j} \Big[C_2\big(g({\cal E}^{+}_j,T)\big) -
C_2\big(g({\cal E}^{-}_j,T)\big)\Big]. \label{eq11}
\end{eqnarray}
Here the summation is taken over chiral edge modes; the integer $j$ counts
chiral edge modes at finite frequency. ${\cal E}^{+}_j$ and ${\cal E}^{-}_j$ stand for
a pair of two energies by which the $j$-th chiral edge mode is bounded (Fig.~\ref{s-fig5}(b)).
We define ${\cal E}^{+}_j > {\cal E}^{-}_j$ when the $j$-th chiral edge mode is
right-handed, while ${\cal E}^{-}_j > {\cal E}^{+}_j$ when the mode is left-handed. %(Fig.~\ref{fig2}(b)).
$C_2(x) \equiv  \int^{x}_0 dt \ln \big((1+t)/t\big)^2$ and $g(E,T)$ is the
Bose distribution function.

The above expression is qualitatively consistent with the thermal magnon Hall
conductivity in the clean limit, which was previously obtained based on the linear response
theory~\cite{s-matsumoto11a,s-matsumoto11b,s-matsumoto14,s-qin11,s-qin12};
\begin{eqnarray}
 \kappa_{xy} = - \frac{k^2_B T}{\hbar} \sum_{n}\int \frac{d^2{\bm k}}{(2\pi)^2}
 \Omega_{n}({\bm k}) \!\ \Big[C_2\big(g({\cal E}_n({\bm k}),T)\big) - \frac{\pi^2}{3} \Big].
 \label{eq12}
\end{eqnarray}
Here ${\cal E}_n({\bm k})$ and $\Omega_{n}({\bm k})$ stand for
the $n$-th magnon energy band and $n$-th band Berry's curvature, respectively.
${\bm k}\equiv (k_x,k_y)$ denotes the two-dimensional crystal momentum. The Chern integer
is defined for each band as an integral of the curvature in the first Brillouin zone (B.Z.);
\begin{eqnarray}
{\rm Ch}(n) \equiv \int_{\rm B.Z.}\!\ \Omega_{n}({\bm k}) \!\ \frac{d^2{\bm k}}{2\pi}. \nonumber
\end{eqnarray}
When all bulk magnon bands are
fully gapped: ${\cal E}_{n}({\bm k}) > 0$ for all ${\bm k}$ and for all $n$,
a sum of the Chern integers over band is zero~\cite{s-shindou13a};
$\sum_{n} {\rm Ch}(n) = 0.$ Thereby, Eq.~(\ref{eq12}) reduces to
\begin{eqnarray}
 \kappa_{xy} = -\frac{k^2_B T}{h} \sum_{n}\int \frac{d^2{\bm k}}{2\pi}
 \!\
 \Omega_{n}({\bm k}) \!\ C_2\big(g\big({\cal E}_n({\bm k}),T\big)\big).
 \label{eq13}
\end{eqnarray}

Eq.~(\ref{eq13}) becomes identical to Eq.~(\ref{eq11}), when
an energy band width of each bulk magnon band %(call as $W_n$ a band width of the $n$-th band)
is much smaller than $k_B T$.
In this limit, ${\cal E}_{n}({\bm k})$ in the right hand side of Eq.~(\ref{eq13})
can be replaced by its band center energy $\overline{{\cal E}_{n}}$;
\begin{align}
\lim_{\Delta \ll k_BT} \kappa_{xy} &= -
\frac{k^2_B T}{h} \sum_{n}\int \frac{d^2{\bm k}}{2\pi}
 \!\
 \Omega_{n}({\bm k}) \!\ C_2\big(g\big(\overline{{\cal E}_n},T\big)\big)
 = - \frac{k^2_B T}{h} \sum_{n} {\rm Ch}(n)  \!\ C_2\big(g\big(\overline{{\cal E}_n},T\big)\big).
 \label{eq14}
\end{align}
The equivalence between Eq.~(\ref{eq11}) and Eq.~(\ref{eq14}) can be
seen with a help of the bulk-edge correspondence~\cite{s-halperin82,s-hatsugai93,s-shindou13a}.
%a comparison between .
For example, consider (i) the $n$-th bulk magnon band whose band
center energy is $\overline{{\cal E}_n}$ and
${\rm Ch}(n) = l$ and $\sum^{n-1}_{j=1}{\rm Ch}(j)=-m$ with $l>m>0$.
The bulk-edge correspondence dictates that
$m$ pieces of right-handed chiral edge modes enter into the bulk band
from below and  $(l-m)$ pieces of left-handed chiral edge modes enter into
the band from above (Fig.~\ref{s-fig5}(c)). In the presence of small disorders,
all the bulk states in the $n$-th band are localized except for the delocalized
states at the band center $\overline{{\cal E}_n}$. Thereby, the delocalized bulk states
at the band center terminate all the chiral edge modes;
$\overline{{\cal E}_n}$ bounds the $m$ pieces of right-handed
chiral edge modes from above and the $(l-m)$ pieces of left-handed chiral edge
modes from below (Fig.~\ref{s-fig5}(d)). 
%\begin{eqnarray}
%\overline{{\cal E}_n} = {\cal E}^{+}_{\cdots }={\cal E}^{+}_{\cdots +1}=\cdots={\cal E}^{+}_{\cdots+m}
%={\cal E}^{+}_{\cdots+m+1}={\cal E}^{+}_{\cdots+m+2}=\cdots={\cal E}^{+}_{\cdots+l}. \label{c1}
%\end{eqnarray}
Let us consider another examples: (ii) the $n$-th band with
${\rm Ch}(n) = l$, $\sum^{n-1}_{j=1}{\rm Ch}(j)=m$ and $l>0$, $m>0$. 
In this case, the correspondence tells that $m$ pieces of left-handed 
chiral edge modes pass by the band center energy $\overline{{\cal E}_n}$, 
while $l$ pieces of left-handed chiral modes are terminated by the delocalized 
bulk states at $\overline{{\cal E}_n}$; $\overline{{\cal E}_n}$ bounds the latter 
$l$ pieces of modes from below (Fig.~\ref{s-fig5}(e)).  By considering 
other cases as well and integrating them together, we can readily rewrite 
Eq.~(\ref{eq14}) into Eq.~(\ref{eq11}) in the small band width limit.
Using $C_2(g(E=0+))=\frac{\pi^2}{3}$, we can further
generalize the argument so far into a case with complete flat zero energy
bands, ${\cal E}_{n}({\bm k})=0$ for all ${\bm k}$ and for $\exists$ $n$, giving
a consistency between Eq.~(\ref{eq11}) and Eq.~(\ref{eq12}) too.

The thermal Hall conductivity can be used to
confirm the presence/absence of topological chiral edge modes in finite frequency
regimes. For example, the thermal magnon Hall conductivity in the high temperature
limit goes to a {\it non-zero} constant value ! Moreover, the value is given by 
a sum of those mobility gaps which bound topological chiral edge modes;
\begin{eqnarray}
%\lim_{T\rightarrow \infty}\kappa_{xy} = \left\{\begin{array}{ll}
%\frac{k_B}{h} \sum_{j} \big({\cal E}^{+}_{j}-{\cal E}^{-}_j\big), &  \!\ \!\ \!\ \ \ \ {\rm with} \!\ \!\
%{\rm disorders}, \\
%\frac{k_B}{\hbar} \sum_{n} \int \frac{d^2{\bm k}}{(2\pi)^2}
%\!\ \Omega_{n}({\bm k}) \!\ {\cal E}_{n}({\bm k}), & \!\ \!\ \!\ \ \ \ {{\rm without}} \!\ \!\ {\rm disorders}.  \\
%\end{array}\right.
\lim_{T\rightarrow \infty}\kappa_{xy} =
\frac{k_B}{h} \sum_{j} \big({\cal E}^{+}_{j}-{\cal E}^{-}_j\big). \label{eq17}
\end{eqnarray}
Here the summation is over the edge modes;
${\cal E}^{+}_{j}$ and ${\cal E}^{-}_j$ bound the $j$-th chiral edge mode in pair.
Note that ${\cal E}^{+}_j>{\cal E}^{-}_j$ for the right handed chiral edge mode
and ${\cal E}^{+}_j<{\cal E}^{-}_j$ for the left handed mode; right/left
handed chiral mode contributes to postive/negative thermal Hall conductivity
respectively. Nonetheless, there is no `topological' reason which requires the
sum in Eq.~(\ref{eq17}) to be zero. The non-zero $\kappa_{xy}$ in the high
temperature limit is quite unconventional. The feature clearly
tells the presence of the chiral edge modes from otherwise
in actual experimental systems.

\end{document}